\magnification=1200

\pageno=1
\centerline {\bf BEYOND STRINGS, MULTIPLE TIMES  AND  GAUGE THEORIES OF  }
\centerline {\bf    AREA-SCALINGS       RELATIVISTIC TRANSFORMATIONS    }
\medskip
\centerline { Carlos Castro }
\centerline { Center for Theoretical Studies of Physical Systems   }
\centerline { Clark Atlanta University}
\centerline {Atlanta, Georgia 30314}
\centerline {}
\smallskip

\centerline {\bf July  1997 }
\medskip
\centerline {\bf ABSTRACT}
\medskip

Nottale's special scale-relativity principle was proposed earlier by the author as a plausible geometrical origin to string theory and extended objects . Scale Relativity is to scales what motion Relativity is to velocities. The universal, absolute, impassible, invariant scale under dilatations in Nature is taken to be the Planck scale,  which is not the same as the string scale. Starting with ordinary actions for strings and other extended objects, we show that gauge theories of  volume-resolutions scale-relativistic  symmetries,  of the world volume measure associated with the extended ``fuzzy'' objects,  are a natural and viable way to formulate the geometrical principle underlying the theory of all extended objects. Gauge invariance can only be implemented if the extendon actions in $D$ target dimensions are embedded in $D+1$ dimensions with an extra temporal variable corresponding to the 
scaling dimension of the original string coordinates. This is achieved upon viewing the extendon 
coordinates, from the fuzzy worldvolume  point of view,   
as noncommuting matrices valued in the Lie algebra of Lorentz-scale relativistic transformations.       
Preliminary steps are taken to merge motion relativity with scale relativity by introducing the gauge field  that gauges the Lorentz-scale symmetries in the same vain that the spin connection gauges ordinary Lorentz transformations and, in this fashion, one may go beyond string theory to construct the sought-after General Theory of Scale-Motion Relativity. Such theory requires the introduction of the scale-graviton ( in addition to the ordinary graviton)  which 
is the field that gauges the symmetry which converts motion dynamics    into 
scaling-resolutions dynamics and vice versa ( the analog of the gravitino that gauges supersymmetry). 
To go beyond the quantum string geometry  most probably would require  
a curved fractal spacetime description ( curved from both scaling and motion 
 points of views) with a curvilinear fractal coordinate system. Non-Archimedean geometry and $p$-adic numbers are 
essential ingredients comprising   
the geometrical arena of such extensions of quantum string geometry.

\centerline {\bf  I. Introduction}

Despite the tremendous progress in string theory in the past two years a geometrical foundation of the theory, like Einstein's General Theory of Relativity,  is still lacking. The fact that string theory 
contains gravity is, per se, no satisfactory explanation of what is the underlying reason behind 
it  nor why it should  be so, especially insofar that a background independent formulation of string theory is still missing.  Spacetime should emerge from the string . Tsetlyn [1] has remarked that the vanishing of the $beta$ functions  for the couplings of the non-linear $\sigma$ model associated with the string propagation in curved backgrounds must contain a clue to the non-perturbative and geometrical 
formulation of string theory.

Some time ago, Amati, Ciafaloni and Veneziano [4] pointed out that some sort of enlarged equivalence principle is operating in string theory in which dynamics is not only independent of coordinate transformations but also of structures ocurring at distances shorter than the string length in units of $c=1$, $\lambda_s =\sqrt { 2\alpha \hbar}$, where distances smaller than $ \lambda_s$ are not relevant in string theory. The string scale differs from the Planck scale $\Lambda =\sqrt { (\hbar G/c^3)}$ where $G$ is Newton's  constant in four dimensions. 
There is however an intermediate regime  between both scales and $D$-branes probes have been suggested to explore such 
distances below $\lambda_s$. Yoneya and references therein  [23].

In [3] we obtained modifications/corrections/extensions to the stringy-uncertainty principle within the framework of the theory of special Scale-Relativity developed  by Nottale [2]. In particular, the size of the string was shown to be 
bounded by the minimum Planck scale and an upper, impassible, absolute  scale, invariant under dilations, that incorporates the principle of scale-relativity to cosmology as well. Such principle has allowed Nottale to propose a very elegant and ``simple'' resolution to the cosmological constant problem. Since the Planck scale plays the same role in scale relativity  that the speed of light did in special relativity,  
the Planck energy $decouples$  from the Planck scale : it takes an $infinite$ amount of energy to probe 
finite Planck scale resolutions ! hence  the Planck scale serves as a natural ultraviolet regulator in QFT. 
Whereas the upper scale serves as an infrared one. 
For the importance that the Planck scale may have as a natural regulator of matter QFT in quantum gravity, 
noncommutative geometry, quantum groups, .... we refer to the authors [5,14,15,25,27] 
and the role that loop spaces, spin networks, graphs and networks,....
may have in the discrete structures of spacetime at Planck scale  we refer to [6,16,17,22].

The aim of this work is to argue that  the principle of scale-relativity can be incorporated into the theory of extended objects [3] ; i.e. we will show that applying this principle to the $resolutions$ 
( that  any physical measurement  can make no smaller than the Planck scale )   of the extended object's  world-volume coordinates, a 
$fuzzy$  $p$-brane, [19], allows the implementation of  the principle of Lorentzian-scale relativistic invariance to the 
world-volume measure of these extended objects,  as they are dynamically embedded in a curved target spacetime background. Scale-relativity invariance of area-resolutions, volume-resolutions,....  
can be maintained in all Dirac-Nambu-Goto actions of extended objects
if one enlarges the $D$ background spacetime   to an effective  $D+1$ spacetime where an additional $temporal$ dimension is added. This extra $time$ dimension stems from the $scaling$ dimension of the original $p$-brane $X^\mu$ 
coordinates. The scaling dimension is determined by the $X^\mu$ coordinates transformation properties 
under volume-resolutions
scale-relativistic transformations. For example, the $10D$ superstring would require an $11D$ space of signature 
$(9,2)$. The $11D$ supermembrane requires a $12D$ space of signature $(10,2)$ etc.....which is what Vafa and Bars have been advocating recently concerning, F, S , theories. [31].

Once area-resolutions, volume-resolutions,....are included in the physical processes, the $X^\mu$ $fuzzy$ coordinates are metamorphosed into $matrices$ carrying internal indices 
associated with the Lie-algebra generators of scale-relativistic transformations of the world-volume resolutions.  In the string case, the 
$X^\mu \rightarrow X^\mu_{\Delta^a \Delta^b }\Sigma^ {\Delta^a \Delta^b }$ where the matrix generators are $\Sigma^ {\Delta^a,\Delta^b }$. In the past months there has been growing evidence that 
the $D=11$ M theory in the infinite momentum frame bears a connection to large $N$ Matrix models ; i.e supersymmetric gauge quantum mechanical models [26] associated with the algebra of area-preserving diffs ( Hoppe 
, Flume, Baake et al) , and that the string coordinates can be viewed as noncommuting matrices resulting from the collective modes of the positions of $D$ branes 
( Witten, Polchinski). The membrane ground state appears  as a ``condensation'' of an infinity of $D$-0-branes.

In our case, the $D~X^\mu$ matrices are then embedded into a larger space
of $D+1$ matrices  $Z^M$, where the extra $temporal$ dimension is the resolution $dependent$ scaling dimension, $\delta (\Delta \sigma^a)$  
of the original $X^\mu$ coordinates ( matrices) .  
After the metric $G_{\mu\nu}$ is embedded into ${\cal G}_{MN}$ we show that the Dirac-Nambu-Goto action of the fuzzy string moving in the enlarged $D+1$ spacetime is in fact invariant under 
area-resolutions scale-relativistic symmetries and, most importantly, 
the scale-relativistic version of the $beta$ functions,   $\beta^{\cal G} _{MN}\not=,\beta^Z \not=0$.    
One can find vanishing $\beta^{\cal G}_{MN},\beta^Z$  but the solutions are $trivial$.  If one wanted to find nontrivial solutions this would entail imposing unnatural constraints  
among the $X^\mu$ coordinates.  
i.e. scale-relativistic 
transformations preserving  the world-volumes of the extended objects are compatible with the 
$\beta^{\cal G} _{MN}\not=0, \beta^{Z^M} \not=0$ conditions. This should be the starting point to explore 
string propagation in non conformal backgrounds. Although we must remark that the scale-relativity $beta$ functions are $not$ the same as the ordinary $2D$ conformal field theory ones.

The main point of this work is that in the same 
way that Lorentz invariance requires a Minkowski spacetime, volume-resolution scale-relativistic invariance requires an extra scaling 
temporal dimension so the effective spacetime is $D+1$ dimensional. Furthermore, instead of restricting ourselves to 
volume-scalings but instead concentrating  on the most general coordinate-resolutions scale-relativistic transformations  without the restriction of scaling the volumes as a $whole$, one will have then $many$ scaling-temporal 
dimensions. This could be an explanation of the multiple-time/signatures  spacetime backgrounds which are 
very popular to-day.

Care must be taken not to confuse transformations involving the resolutions of the fuzzy worldsheet coordinates with the string worldsheet coordinates themselves !.   
Extended objects can also  be interpreted as  a gauge theory of volume preserving diffeomorphisms [11,12]. 
 We were able to  show in [11] that $p$-branes can be seen  as composite antisymmetric tensor field theories  of 
the volume preserving diffs ( of the type proposed by 
Guendelman, Nissimov and Pacheva [11]   and using the local gauge theory reformulation of extended objects 
developed by 
Aurilia, Ansoldi, Smailagic and Spallucci [11] ,). 
The relevance of these composite theories is that it allowed the author to build-in , from the very beginning and 
$without$ any conjectures,  the analogs of $S$ and $T$ duality symmetries in extended objects. Furthermore, we also 
have shown  
that the 
$p$-branes worldvolumes  had a natural correspondence with nonabelian composite-antisymmetric tensor fields 
which were $not$  of the Yang-Mills type fields, ( only for the membrane case ) , verifying the conjecture of 
Bergshoeff et al [11].

After reviewing Nottale's essential results in section  II we 
proceed  into the implementation program of scale-relativity in {III and show that strings can be viewed 
as gauge theories of area-resolutions scale-relativistic transformations in the enlarged 
$D+1$ spacetime including the extra temporal scaling variable. Also in III 
we discuss  how one might achieve the goal of building the general theory of scale-motion relativity and go 
beyond string theory. 

Finally, in the conclusion, other topics are briefly discussed, like  Weyl-Finsler geometries, 
renormalization group flows, curved backgrounds, 
extensions of the quantum string geometry [33] based on curved fractal spacetimes, non-Archimedean geometry and $p$-adics numbers, among some.  The  
supersymmetrization program should be carried out to see how far these ideas can be taken and be able to make comparisons with the current results of string-duality, moduli spaces, etc..... 

\bigskip
\centerline {\bf II.  }
\bigskip

We shall present  a brief review of Nottale's principle of special scale relativity. For a detailed account of the theory of Scale Relativity that orginated with the study of fractals , we recommend the reader to study  Nottale's  work that appeared in [2]. The 
principle of coordinate-resolutions special scale-relativity is essentially stated : 
``  that the laws of physics should be covariant  
under any state of `` scaling-motion'' of all frames  of reference  associated with all systems  of observers
( carrying coordinate-resolutions rulers  
with their physical measurements apparatus ). The Planck scale is taken to be the minimum-resolution scale in Nature.

Essentially, one has a collection of scalar fields, $\varphi$,  that 
under Lorentzian-scalings (  of the spacetime coordinate $resolutions$  
that our physical apparatus can resolve ) behave like the ordinary spacetime coordinates under ordinary Lorentzian transformations ( change of frame of reference).  The analog of the speed of light, $c$, is played by the logarithm of the $ratio$ of two resolutions : One is the resolution, $\lambda_o$,  with respect to which we measure other resolutions, $\Delta x_o\leq \lambda_o$; and another is the Planck scale, $\Lambda$ in the appropriate dimension. The analog of time is played by the scaling dimension of the scalar fields, $\varphi$, which Nottale labeled by $\delta$. The origins of scale-relativity were motivated by the fractality of spacetime microphysics and, the very plausible fractality of cosmological structures as well. Ord [2], found that the relativistic quantum mechanics of particles could be reinterpreted in terms of fractal trajectories : continuous but nowhere differentiable in spacetime. 

We do not intend to fall hostage  of the debate of ``what is quantum'' and what is ``classical''. Nottale's view is that a nowhere differentiable spacetime should not be viewed any longer as ``classical''. Our aims are less ambitious. We just go ahead and ask ourselves whether or not one can implement the scale-relativity principle to string theory and extended objects. We start with some definitions :         

The scaling behaviour of $\varphi=\varphi (x,\Delta x)$ under scale-relativistic transformations  is :

$$ \Delta x_0\rightarrow \Delta x.~~~ln(\varphi /\varphi_o)=ln [(\Delta x_0/\Delta x)^{\delta (\Delta x)}] 
\Rightarrow \varphi (x,\Delta x)=\varphi_o (x)(\Delta x_0/\Delta x)^{\delta (\Delta x)}.\eqno (2.1)$$
The scale dimension of the field $\varphi (x, \Delta x) $ is measured w.r.t the frame of reference whose resolution is 
$\Lambda \leq \Delta x_0 \leq \lambda_o$ . Since the quantity $\lambda_0/\Lambda$ will play the role of the speed 
of light, $c$, and because $\Lambda$ is taken to be the invariant scale this implies also that 
$\lambda_o$ must be  a fiducial and $fixed$ scale w.r.t which we measure the running scales. This $\lambda_o$ scale was taken by Nottale to be the scale which signals the classical geometry-fractal spacetime transition and usually is taken to be the deBroglie wavelength of electron, for example. However, this scale is free. One may set it equal to the string scale if one wishes. The scaling dimension is then :   

$$\delta (\Delta x) ={\delta_o (\Delta x_0) \over \sqrt {1-{  ln^2 (\Delta x_0/\Delta x)\over ln^2(\lambda_0/\Lambda)}}},~~~   \Lambda \leq        \Delta x_0 \leq \lambda_0. \eqno (2.2)$$
 
Under two consecutive Lorentz-scale transformations of the $resolutions$ : $\Delta x_0 \rightarrow \Delta x$
and $\Delta x \rightarrow \Delta x'$  , the  logarithm of $\varphi$ and the scaling dimension $\delta$ transform like the components of a two-vector :

$$\Delta x_0 \rightarrow  \Delta x \rightarrow \Delta x'.~~~ ln(\varphi'/\varphi_o)={ln(\varphi/\varphi_o) -\delta ln \rho \over 
 \sqrt {1-{  ln^2\rho \over ln^2(\lambda_0/\Lambda)}}}.\eqno (2.3)$$

$$\delta'=
   { \delta+  { (ln\rho)( ln {\varphi\over \varphi_0})\over 
ln^2(\lambda_0/\Lambda) }
\over \sqrt {1-{ln^2 \rho \over ln^2(\lambda_0/\Lambda)}}}.
\eqno (2.4 )$$
 
where the composition of dilations ( analog of addition of velocities ) is :

$$ln \rho ={ln(\Delta x /\Delta x_0)\pm ln(\Delta x'/\Delta x)\over 
1\pm {ln (\Delta x /\Delta x_0 )ln(\Delta x'/\Delta x)\over ln^2(\lambda_0/\Lambda)}}.
\eqno (2.5)$$

If one chooses, $\Delta x_0 =\Delta x $ then in this particular case one recovers  
$\varphi (x,\Delta x_0)\equiv \varphi_o (x)$ so the above equations simplify :

$$ln \rho =ln (\Delta x'/\Delta x)=ln (\Delta x'/\Delta x_0   ,~~~\varphi'(x,\Delta x') =\varphi_o (x) (\Delta x'/\Delta x)^{-\delta'}.\eqno (2.6)$$
where :

$$\delta ' =
{\delta (\Delta x_0) \over \sqrt {1-{ln^2 \rho \over ln^2(\lambda_0/\Lambda)}}}
=\delta (\Delta x_0) [(1-\beta^2)^{-1/2}].
\eqno (2.7)$$                
$\beta \equiv 
[ln(\rho)/ln(\lambda_o/\Lambda)]$ which is the analog of $v/c$ in motion relativity, one can recognize eq-(2.7) as the analog of time dilation in motion-relativity.

We could  set 
$c=ln(\lambda_o/\Lambda)=1$. 
Eq-(2.6) has exactly the $same$ form as (2.1), as it should if covariance is to be maintained. Henceforth, we shall omit the suffix $\Delta x'$. A finite Lorentz-scale transformation from the fiducial scale-frame of reference  
$\Lambda \leq \Delta x_0 \leq \lambda_o, $ to a new scale $\Delta x$ implies :

$$ \Delta x_0 \rightarrow \Delta x,
~\varphi_o (x,\lambda_o)\rightarrow \varphi_o (x)e^{-\beta \delta }=\varphi (x,\Delta x),
~\delta_o (\Delta x_0)\rightarrow \delta_o (\Delta x_0) (1-\beta^2)^{-1/2}=$$
$$  \delta (\Delta x).~~~       \beta={ln (\Delta x/\Delta x_0)\over ln(\lambda_o /\Lambda)}. \eqno (2.8)$$ 

The infinitesimal scaling transformations from one frame, whose relative velocity w.r.t the 
fiducial frame 
is $\beta$,  to another frame whose relative velocity w.r.t the fiducial frame is 
$\beta +\Delta \beta$ 
are obtained  from the relations  :

$$ln (\varphi/\varphi_o)=-\beta \delta.~~ ln (\varphi'/\varphi_o)=-\beta' \delta'. ~~\beta'=\beta +\Delta \beta.~~~\delta (\beta +\Delta \beta)=\delta +\Delta\beta {\partial( \delta)\over \partial \beta} .\eqno ( 2.9)$$
so under infinitesimal scaling-relativistic transformations :

$$\delta_\beta [ln (\varphi/\varphi_o)]=ln (\varphi'/\varphi_o)-ln (\varphi/\varphi_o)= 
- [(\beta +\Delta \beta)(\delta +\Delta \delta)-  \beta \delta]\sim -(\gamma^2 \delta) \Delta \beta.
\eqno (2.10)$$ 
Therefore, we shall define the  infinitesimal scaling-relativistic transformations :

$$\delta_\beta  [ln (\varphi/\varphi_o)]=     
{\delta ln (\varphi/\varphi_o)\over \delta \beta}\Delta \beta =-(\gamma^2 \delta)  \Delta \beta 
\Rightarrow \delta_\beta  [ (\varphi/\varphi_o)]=  -(\gamma^2 \delta)[(\varphi/\varphi_o)] \Delta \beta . 
\eqno (2.11) $$
where we explicitly write $[{\delta ln (\varphi/\varphi_o)/\delta \beta}]$ to emphasize the 
fact that one is 
performing a resolution-scaling transformation, a change of the scaling frame of reference,  and 
$not$ a differentiation w.r.t the $\beta$ variable. In ordinary relativity we don't have expressions like $\partial X/\partial v$ where $v$ is the relative velocity between two frames of reference.  
The latter equations  show that $\delta,ln(\varphi/\varphi_o),\beta$ play the same role as time, space coordinates and velocity, respectively,  in motion relativity. The scaling-dimension ( a function of the 
resolutions) associated with   the field $\varphi$, 
$\delta(\varphi)$,  is 
evaluated at two different points, $\beta, \beta+\Delta \beta$;  however the functional form of $\delta$ does not 
change.   
Notice the subtlety in the difference upon naive differentiation w.r.t the $\Delta x $ and 
performing a scaling transformation. The reason is the following.

Extreme caution must be 
taken in order not
to  confuse the scaling dimension, $\delta$, with its $transformation$ properties under Lorentz-scalings. 
For example, the quantities $\Delta x' ,\Delta x,\Delta x_0      $ can all $flow$ in such a fashion that their respective $ratios$ 
remains constant. Imagine scaling the  $\Delta x' ,\Delta x ,  \Delta x_0      $ scales  by a common factor 
( the $\lambda_o$ and $\Lambda$ scales remain $fixed$ in all the  formulae ) 
so their $ratio$ remains invariant,  then the quantity    
$\beta =[ln(\rho)/ln(\lambda_o /\Lambda)] $ still remains constant since both $v,c$ do.   
This means that the gamma-dilation factor does not change either. We can notice 
also that 
the scaling velocity $\beta=v/c$ and the gamma dilation factor is also invariant under the analog of a 
T duality transformation $R\leftrightarrow (1/R)$: 
$\Delta x/\Delta x' \rightarrow (\Delta x'/\Delta x)$ and $(\lambda_o /\Lambda)\rightarrow 
(\Lambda /\lambda_o )$.

However, the quantity $\delta (\Delta x)$ does $change$ because $\Delta x$ has $flowed$.
A flowing value for $\Delta x$ is not the same as a change of a reference frame . One must not confuse the values that a coordinate, in a given frame ,   can take with its $transformation$ properties under Lorentz-scale transformations. 
We shall take $c=1$ from now on and by choosing a frame of reference we mean fixing the value of the relative velocities $v/c=v=\beta=ln (\rho)$ in (2.5,2.6) despite the fact that both quantities $\Delta x, \Delta x'$ can both flow maintaining its ratio fixed. It is in this context that there is a crucial difference between 
taking ordinary differentiation of $\varphi (x, \Delta x)$ w.r.t the $\Delta x$ flowing variable  and 
performing an infinitesimal scaling transformation,  $\delta_\beta \varphi$, given by eq-(2.10,2.11). 
Therefore, by a scaling transformation one means :

$$  \delta_\beta    ln (\varphi/\varphi_o)      =  \Delta \beta {\delta ln (\varphi/\varphi_o)\over \delta \beta} =-\gamma^2\delta \Delta \beta.\eqno (2.12) $$
And by the analog of ``scaling-motion'' (resolution-motion)  of the  $\varphi$ field 
w.r.t the   
fiducial reference field, $\varphi_o$, 
due to the $flowing$ values of $\Delta x$ , ( imagine the motion of a particle whose coordinate 
is $\varphi$ moving with velocity $\beta$ w.r.t a    
particle  whose  frame of reference carries the coordinates $\varphi_o$)  :

$$ ln (\varphi^2 (\Delta x) /\varphi_o)-
ln (\varphi^1 (\Delta x)/\varphi_o)= ln (\varphi (\Delta x^2) /\varphi_o)-
ln (\varphi (\Delta x^1)/\varphi_o).\eqno (2.13)$$
and the scale-velocity is : 
$${d [ln (\varphi/\varphi_o)] \over d \delta}|_\beta =-\beta. \eqno (2.14) $$ 
where we implement the resolution-motion of the ``coordinates'' $\varphi$ between two instants of time, 
$\delta^1, \delta^2$, by defining two functions,  $\varphi^1(\Delta x), \varphi^2(\Delta x)$, obeying :

$$\varphi^1 (\Delta x)=\varphi (\Delta x^1).~~~\varphi^2 (\Delta x)=\varphi (\Delta x^2)$$
$$\delta (\varphi^1)= \delta^1 (\Delta x)=\delta (\Delta x^1).~~~\delta (\varphi^2)=  \delta^2 (\Delta x)=
\delta (\Delta x^2). \eqno (2.15)$$
It is now when we can speak of a scaling-coordinates interval, $\Delta (\varphi/\varphi_o)$, versus  a 
scaling-time interval, 
$\Delta \delta$, w.r.t a given fiducial frame of reference. 
In this fashion it is sensible  to view the scaling dimension as the true analog of a time coordinate. 
This is not new in string theory and quantum mechanics/quantum cosmology. The Liouville mode in non-critical strings has played  the role of a ``temporal'' direction as advocated many times by the authors [10] in connection to the origin of the arrow of time. The quantum phase space origins of a point particle from string solitons and $D$ brane scaffolding dynamics can also be studied within this framework [10]. The importance that duality and scaling in quantum mechanics has in connection with the emergence of an intrinsic fractal scaling time variable 
was discussed by Datta [18].

Summarizing, 
the analog of a time interval w.r.t a  fiducial  frame of reference in scale relativity is :

$$\Delta (\delta) \equiv \delta^2 -\delta^1=\delta (\Delta x^2)-
\delta (\Delta x^1);~
\Delta (\delta ') =\delta '^2 -\delta '^1=
\gamma \Delta (\delta);~\gamma =[1-\beta^2]^{-1/2}.\eqno (2.16)$$
and the  relative ``velocity'' between the $\Delta x', \Delta x $ frames    
 is the one given in the r.h.s of (2.5).         
Similar reasoning applies to the analog of a spatial interval, $\Delta \varphi_=\varphi^2 -\varphi^1$, where $ \varphi^2=\varphi (\Delta x^2);\varphi^1=\varphi (\Delta x^1)$ ,    

Therefore,  the 
scale-relativistic    analog of a Lorentz invariant spacetime world line interval  is :

$$ d\eta ^2 =[ln^2 (\lambda_0/\Lambda)] (d\delta)^2 -{ (d  (\varphi/\varphi_o ))^2 
\over (\varphi/\varphi_o)^2}= [ln^2 (\lambda_0/\Lambda)]
(d\delta)^2[1  - { 1\over (ln^2 (\lambda_0/\Lambda))} { (dln(\varphi/ \varphi_o )^2 \over (d\delta)^2 }]. \eqno (2.17)$$
and one gets the usual time dilation expression : $\gamma (d\eta)=d\delta$
where we have used in the last term of (2.17) the scaling-velocity relation :

$$ln(\varphi/\varphi_o)=-\beta \delta \Rightarrow {dln(\varphi/\varphi_o)\over d \delta}|_\beta 
=-\beta.
\eqno (2.18)$$
  
Therefore,  in all frames we have the scale-relativistic invariant analog of proper time :

$${(d\delta)^2 \over \gamma^2}={(d\delta')^2 \over \gamma'^2}={(d\delta'')^2 \over \gamma''^2}=
.....={(d\delta_o)^2 \over \gamma_o^2}={(d\delta_o)^2 \over 1}=(d\eta)^2. \eqno (2.19)$$

For a collection of fields , $\varphi^i$,  $all$ with the $same$ scaling dimensions one has the 
generalization of flat Minkowski spacetime :
$$ d\eta^2 =[ln^2 (\lambda_0/\Lambda)] (d\delta)^2 -{ \sum_i (d\varphi^i)^2 
\over \sum_i(\varphi^i)^2}.\eqno (2.20)$$
from now on we shall  omit the $\varphi_o$ in our formulae for convenience but should not be forgotten !

The two-dim metric in (2.11) is flat : $dT^2-dU^2$ with $U=ln(\varphi)$. Similarly, the metric in (2.20) is also flat as one can see by performing the suitable  change of coordinates  :

$$ { \sum_i (d\varphi^i)^2 
\over \sum_i(\varphi^i)^2}=
\sum_i { (d\varphi^i)^2 
\over \sum_i(\varphi^i)^2}=\sum_i [d(ln \zeta ^i)]^2 =\sum_i (dU^i)^2.$$
$$U^i= ln(\zeta^i) =\int d[ln(\zeta ^i)]= \int {d\varphi^i \over \sqrt { \sum_i (\varphi^i)^2}}.~i=1,2,3....\eqno (2.21 )$$
Setting $ln^2 (\lambda_0/\Lambda)=1$, the interval becomes  : 

$$ d\eta^2 =[ln^2 (\lambda_0/\Lambda)] (d\delta)^2 -{ \sum_i (d\varphi^i)^2 
\over \sum_i(\varphi^i)^2}=  (d\delta)^2-  \sum_i (dU^i)^2. \eqno (2.22)$$
and it is invariant under scale-relativistic transformations. The interval (2.22) is the analog of the spacetime interval in Minkowski space of signature $(+,-,-,-,...)$. This ``completes'' the review of Nottale's scale relativity.

\bigskip

\centerline{ \bf III. Strings as Gauge Theories of Area Scaling-Relativistic Transformations }

\medskip
\centerline{ \bf 3.1 Area-Scale-Relativity }

In analogy with the transformations given in the previous section by Nottale we can  
define the scalings under area-resolutions where the Planck area, $\Lambda^2$, is chosen to be the minimum resolution of area in nature. Now we define :

$$\varphi (x^1,x^2;\Delta x^1, \Delta x^2)\equiv   \varphi (x^1,x^2;\Delta x^1\wedge \Delta x^2)=
\varphi (x^1,x^2;\Delta A). \eqno (3.1)$$

Mutatis mutandis 

$$\Delta A=\Delta x^1 \wedge \Delta x^2 \rightarrow  
\Delta A'=\Delta x'^1 \wedge \Delta x'^2 =\rho \Delta A. \eqno (3.2a)$$ :

$$ ln(\varphi'/\varphi_o)={ln(\varphi/\varphi_o) -\delta (ln \rho) \over 
 \sqrt {1-{  ln^2\rho \over ln^2(   \lambda_o^2        /\Lambda^2)}}}.~~~   \lambda_o^2  \geq  \Delta A_0 \geq   \Lambda^2     \eqno (3.2b)$$

$$\delta'=
   { \delta+  { (ln\rho)( ln {\varphi\over \varphi_0})\over 
ln^2(  \lambda_o^2              /\Lambda^2) }
\over \sqrt {1-{ln^2 \rho \over ln^2(      \lambda_o^2           /\Lambda^2)}}}.
\eqno (3.3)$$
 where the composition of area dilatations ( analog of addition of velocities ) is :

$$ln \rho ={ln(\Delta A /\Delta A_0)\pm ln(\Delta A'/\Delta A)\over 
1\pm {ln (\Delta A/\Delta A_0)ln(\Delta A'/\Delta A)\over ln^2(    \lambda_o^2              /\Lambda^2)}}.
\eqno (3.4)$$

A finite area-resolution Lorentz-scale transformation implies :

$$ \Delta A_0 \rightarrow \Delta A,
~\varphi_o (x,\Delta A_0)\rightarrow \varphi_o (x,\Delta A_0)e^{-\beta  \delta },
~\delta (\Delta A_0)\rightarrow \delta (\Delta A_0) (1-\beta^2)^{-1/2}.  \eqno (3.5)$$

$$\beta={ln (\Delta A/\Delta A_0)\over ln(     \lambda_o^2              /\Lambda^2)}. $$ 
Identical results occur for $p$-branes when volume-resolutions scaling-relativistic transformations 
replace area-scalings.  

\medskip

\centerline {\bf 3.2 Area-Scale-Relativity and Strings}

\medskip

We are now ready to implement the scale-relativistic transformation to string theory and extended objects; i.e. to the 
Nambu-Goto actions.  Lets take the string case as example. The Dirac-Nambu-Goto action :

$$S=\int d\sigma^1d\sigma^2 \sqrt { det~|G_{\mu\nu}\partial_{\sigma_1}X^\mu \partial_{\sigma_2}X^\nu |};.  \eqno (3.6)$$
where $G_{\mu\nu}[(X^\mu(\sigma^1,\sigma^2)]$ is the target spacetime metric. 
Our purpose is to embed the $X^\mu$ coordinates into a larger space whose generalized coordinates are $Z^\mu$  and write now  : 
$Z^\mu(\sigma^1,\sigma^2,\Delta \sigma^1,\Delta \sigma^2)$ to denote the resolution dependence as well. 
Similar arguments apply to the superspace formulation of supergravity/supersymmetry where the bosonic coordinates are part of a larger space.   
The target spacetime coordinates are scalar fields from the world sheet point of view. There will be two main obstacles to overcome.

The first one is the following. If the $X^\mu$ are to play similar  role as the previous scalars $\varphi^i$ with common scaling dimension, $\delta$ there will be difficulties in matching the coordinates with the $\varphi,\delta$. Because now there are $two$ independent 
resolutions,  $\Delta \sigma^1,\Delta \sigma^2$ the analog of velocity and scaling dimension will be  :

$$\beta_1 ={ln (\Delta \sigma^1/\Delta \sigma^1_o)\over ln (\lambda^1_o/\Lambda)}.~~~ 
\beta_2 ={ln (\Delta \sigma^2/\Delta \sigma ^2_o)\over ln (\lambda^2_o/\Lambda)}.$$
$$\delta_1 =(1-\beta^2_1)^{-1/2}\delta_1 (\Delta \sigma ^1_o).~~~\delta_2 =(1-\beta^2_2)^{-1/2}\delta_2 (\Delta \sigma 
^2_o). \eqno (3.7)$$
where $(\lambda^1_o, \lambda^2_o)$ are the two reference scales with respect to which we measure the resolutions 
$\Delta \sigma^1,\Delta \sigma^2$, respectively. As such there are two independent scaling dimensions, $\delta_1,\delta_2$ and the two dimensional version of scaling transformations under 
$    \Delta \sigma^1_o \rightarrow \Delta \sigma^1,~\Delta \sigma ^2_o \rightarrow \Delta \sigma^2 $, are

$$\varphi_o (\sigma^1, \sigma^2,  \Delta \sigma^1_o,   \Delta \sigma ^2_ o )\rightarrow \varphi_o (\sigma^1, \sigma^2, 
\Delta \sigma^1_o,   \Delta \sigma ^2_ o)e^{-\beta_1  \delta_1 -\beta_2  \delta_2 }=
\varphi'( \sigma^1, \sigma^2,  \Delta \sigma^1,\Delta \sigma^2). \eqno (3.8) $$

 A problem arises if one wanted to match the $X^\mu$ coordinates with the $\varphi^i, \delta_1, \delta_2$ quantities
because there are now $two$ scaling dimensions but one temporal coordinate $X^0$. We won't delve into the 
possiblity of choosing two temporal dimensions.  Spacetimes with different signatures have appeared recently 
in  the string literature, in Vafa's F theory , and Bars' S theory,  where spacetimes with $D=(10,2),(11,3),..$ involving the propagation of extended objects with signatures $(p,p)$ must be incorported to implement the duality symmetries associated with nonperturbative superstring theories . Nevertheless this could be a possible avenue to pursue and we will make some comments about this below. It is for this reason that it is more natural to study 
the scale relativity principle applied to $areas$ instead of lengths within the context of string theory. $p$-branes 
will require $p+1$ volume scalings and gauge theories of volume-scale relativity. We should not confuse, once again, resolutions with ordinary worldvolume coordinates and gauge theories of volume preserving diffs with 
volume-scale relativity of resolutions.  In this fashion one has $one$ scaling dimension instead of two and then we could 
match the $X^0$ with $\delta (\Delta A_0)$ transforming under area scalings as : 
                                           
$$~\delta (\Delta A_0)\rightarrow \delta (\Delta A_0) (1-\beta^2)^{-1/2},~~~  
\beta ={ln (\Delta A/\Delta A_0)\over ln (\lambda^2_o/\Lambda^2)}.
\eqno (3.9)$$

The second obstacle is that now one should incorporate motion and scaling dynamics on equal footing. 
At this point the scaling dynamics has been trivial ( gauge degrees of freedom). 
If one wished to generalize matters, one must  incorporate the resolution scaling dynamics 
and extend the 
notion of a metric to the resolution ``displacements'' of the type $d(\Delta \sigma^1), ...$
; i.e. for intervals in the world sheet like :

$$ h_{ \Delta \sigma^a  \Delta \sigma^b}d(\Delta \sigma^a) d(\Delta \sigma^b),
h_{ \Delta \sigma^a \sigma^b}d(\Delta \sigma^a) d(\sigma^b),......\eqno (3.10)$$
and write the generalization of the Dirac Nambu Goto action accordingly .

For example, supergravity can be visualized  as the extension of ordinary Riemannian geometry to a supermanifold where the metric gauges translations and the gravitino gauges supersymmetry. Fields are now quantities depending on the 
superspace coordinates $(x,\theta)$ where $\theta$ are the usual Grassmannian valued coordinates. The super Poincare 
group acts as transformations ( translations, rotations)  in superspace. A superspace metric and measure exists where 
the supervielbein has bosonic/fermionic entries ( directions). In this same fashion we must treat the displacement of 
resolutions and its metric. The fields $\varphi (\sigma, \Delta \sigma )$ must be viewed in the same vain as the 
superfield $\Phi (x,\theta)$ with the difference that the resolutions are also bosonic variables. The 
fuzzy string action  
will now involve a generalized area in the extended space comprising coordinates and resolutions.

The fact that translations of an  object can induced scalings in its size was formulated by Weyl himself using his field of dilatations. Nottale has made some interesting remarks  in relation to the electric charge quantization and charge/mass ratios [2] as results of scale-relativistic dilatation gauge invariance. 
Conversely, internal symmetries like strong interactions can induced spacetime diffs has been shown by Ne'eman and Sijacki [7] in their version of chromo-gravity. To present a rough illustration of what is needed to merge 
scalings and motions into a single theory that we may label $omega$ symmetry we will choose a 
four-dim ``fuzzy''  worldvolume whose coordinates are $\sigma^A, \Delta \sigma^A$ 
$A,B=1,2,3,4$. Its tangent space indices are labeled by lower case latin letters $a,b =1,2,3,4$. The string coordinates are just scalars  living on the worldvolume. 
The ordinary vielbein ( graviton) has a correspondence  with    the scale-graviton.   : 

$$e^a_A \partial_a  \rightarrow {\tilde e} ^{\Delta^a}_{\Delta^A}  \partial_{  \Delta^A}.     \eqno (3.12a)$$ 
 
The spin connection $\leftrightarrow$ the scale-spin connection : 
$$\omega ^{ab} _A \rightarrow {\tilde \omega} ^{\Delta^a \Delta^b }_{\Delta^A}  \Sigma_ {\Delta^a \Delta^b }. \eqno (3.12b) $$ 
 and so forth.
The $X^\mu$ string coordinates will behave like matter fields 
( sections of a bundle) and their partners ( analogs of fermionic superpartners) will be the $\Psi^\mu$ fields. Auxiliary fields would be needed in order to match degrees of freedom . Lets call such symmetry that converts 
coordinates into resolutions , the $omega$ symmetry.  Ordinary covariant derivatives require the connections :

$$D_A X^\mu =(\partial_A +\omega ^{ab} _A +{\tilde \omega} ^{\Delta^a \Delta^b }_{\Delta^A})X^\mu.....\eqno (3.11c)$$
and the analogs of curvature/field strengths and torsion quantities would be :

$$R\sim D\omega +\omega \wedge \omega .~~{\tilde R}\sim D{\tilde \omega}  +{\tilde \omega}  \wedge {\tilde \omega}.~~~ T \sim De +\omega \wedge e .~~~{\tilde T} \sim D{\tilde e}  +{\tilde \omega}  \wedge {\tilde e}+....\eqno (3.11d)$$
The actions are of the matter +geometry form :

$$( D_A X^\mu)^2+....+R+{\tilde R} +R^2 + {\tilde R}^2 +torsion +...... \eqno (3.12)$$
Instead of pursuing this approach at this moment we will opt to take the simplest of all scenarios below. 

\bigskip

\centerline {\bf 3.3 Gauge Theories of Area-Scale Relativistic Transformations}
\bigskip
An alternative simpler route than the previous one of constructing the General Theory of Scale-Motion Relativity 
is to work in a ``flat'' background from the $scalings$  point of view. We shall freeze the scaling dynamics by rendering them trivial in the sense that we will set the scaling-geometrical fields to their corresponding flat values; i.e 
the scale-spin connection ${\tilde  \omega}=0$, the scale-graviton     ${\tilde e}$  will be set to its flat value 
( imagine setting the gravitino to a constant multiple of the Dirac gamma matrices) , the 
scaling-field strenghts associated with the scaling-spin connection are zero etc... 
and we recur to the ordinary principle of gauge invariance. We shall incorporate the 
area-resolutions scale-relativistic transformations as an integral part of an internal space where the 
fuzziness  of the string coordinates manifest themselves.    
We proceed first by working with  $D$-dim target spacetime background  and by matching the  
the $X^\mu (\sigma^a)$ string variables, $\mu=0,1,2,....D-1$ with the the original  
$\varphi^i (\sigma^a, \Delta \sigma^a)$ fields.
Where  
$i=0,1,2,....D-1$. Since $X^0$ is a time coordinate one should Wick-rotate it to match the Euclidean form of the 
$\varphi^0$. It is not necessary to  impose   a factorization condition on the $\varphi^i$ fields : 
$ \varphi^i (\sigma^a) \phi^i( \Delta \sigma^a )$ but instead one must view the $X^\mu$ coordinates as 
matrix-valued :

$$\varphi^i ( \sigma^a, \Delta \sigma^a)\leftrightarrow X^\mu (\sigma^a) 
\equiv   \Sigma^ {\Delta \sigma^a \Delta \sigma^b} X^\mu _{\Delta \sigma^a \Delta \sigma^b} (\sigma^a). \eqno (3.13) $$
where  $\Sigma^ {\Delta \sigma^a \Delta \sigma^b}$ are the generators of scale-relativistic transformations.
These involve scaling-rotations and scaling-boosts and don't differ from the usual Lorentz generators of the Lorentz group as we saw in section II. In the string case, these are : two resolutions scaling-boosts and one 
$U(1)$-like  rotation giving a total of $3$ generators. In the area-scaling case there will be one 
scaling-boost only and one rotation giving a total of two generators. 
In the four-dim worldvolume case the counting goes :
$4$ scaling-boosts, and $6$ rotations giving a total of $10$ generators, etc....  
In this fashion the resolution-dependence is encoded in the matrix-generators  indices and the gauge 
transformation of the $X^\mu$ matrices  associated to the area-scale-relativistic transformations is then :

$$  X^\mu_o\rightarrow  X^\mu (\sigma^a) =  X^\mu_o (\sigma^a) e^{-\beta (\Delta A)\delta (\Delta A)} \eqno (3.14)$$
The matrices $X^\mu$ simply ``rotate''  under resolution-scalings as matter fields in gauge theories do. 
 They behave like field strenghts in ordinary gauge 
theories. In order to evaluate $(X^\mu)^2$ requires taking the trace w.r.t the matrix generators indices  
$\Sigma^ {\Delta ^a \Delta ^b}$
( and not an integration w.r.t the internal space fiber coordinates, the $\Delta \sigma^a$ space ). 
The matrix displacement $dX^\mu$ involves taking derivatives w.r.t the 
$\sigma^a$ variables and not w.r.t the $\Delta \sigma^a$.      

Since the scaling dimension $\delta (\Delta A)$ still remains we must add an additional  scaling $time$ 
variable 
denoted  by $T (\Delta A)$ which solely depends on the area-resolutions. 
The new time variable can be thought of as  a multiple of the constant unit matrix where the proportionality factor is a function of $ \Delta \sigma^a; \Delta A$ variables only . This is the analog of the spinorial 
representation of the spacetime coordinates in twistor methods using Pauli spin matrices : 
$X^\mu \leftrightarrow X^0 {\bf 1} +X^i\sigma_i$ . It is no surprising to find similarities between Penrose's 
description of twistors  because in twistor space a point in complexified and compactified spacetime is 
smeared out (fuzzy)  into a complex line in proyective space. 
It is now when one embeds the 
$D$ dimensional  space into a $D+1$  space, $Z^M$, where the common 
scaling dimension of the $\varphi^i$ fields plays the role of the additional $time$  coordinate. The 
scale-relativistic invariant world interval  is then equated to ( we have ommitted the $\varphi_o$ )  :

$$ (d\eta)^2= {\cal G}_{MN}dZ^M dZ^N =dT^2+G_{\mu\nu}dX^\mu dX^\nu \leftrightarrow (d\delta)^2 -{\sum_i (d\varphi^i)^2 \over \sum _i (\varphi^i)^2}. \eqno (3.15)$$

$$dX^\mu dX^\nu =tr [\Sigma^ {\Delta ^a \Delta ^b} dX^\mu _{\Delta ^a \Delta ^b}(\sigma^a)  
\Sigma^ {\Delta ^c \Delta ^d} dX^\nu _{\Delta ^c \Delta^d} (\sigma^a)]. $$

$$dX^\mu _{\Delta ^a \Delta^b} (\sigma^a)={\partial X^\mu _{\Delta ^a \Delta^b} (\sigma^a)\over \partial \sigma^a}d\sigma^a. $$

$$ G_{\mu\nu}dX^\mu dX^\nu \leftrightarrow - {\sum_i (d\varphi^i)^2 \over \sum _i (\varphi^i)^2}. ~~~
X^\mu \leftrightarrow    \int  {(dZ^m) \over \sqrt { \sum_n (Z^n)^2}}  \leftrightarrow \int  {(d\varphi^i) \over \sqrt { \sum_i (\varphi^i)^2}}. \eqno (3.16) $$
where $iX^0\leftrightarrow \varphi^0 \leftrightarrow Z^0$. 

The ordinary string action is :   
$$S= \int d^2\sigma \sqrt { det[h_{ab}]}.~~~h_{ab} =G_{\mu\nu}\partial_a X^\mu \partial _b X^\nu. \eqno (3.17) $$
with $h_{ab}$ being  the induced worldsheet metric due to the string's embedding in the ordinary 
spacetime.    
Adding the scaling dimension as the extra $time$ dimension yields the extended action  :

$${\cal S} = \int d^2\sigma \sqrt { det[H_{ab}]}.~~~H_{ab} ={\cal G}_{MN}\partial_a Z^M \partial _b Z^N.\eqno (3.18) $$
and now $H_{ab}$  is the induced world sheet metric due to the string's embedding into  the extended space 
$Z^M=T,X^\mu$ space of dimension $D+1$. 
Imposing invariance of the extended-action ${\cal S}$ under area-scaling-relativistic transformations :

$$\delta _\beta {\cal S}=0 \Rightarrow {1\over 2}\sqrt { det[H_{ab}]}H^{ab} \delta_\beta H_{ab}=0. \eqno (3.19)$$

Eq-(3.19 ) is trivially satisfied as a result of the definition of the induced world sheet metric. It is fairly clear  that if one had started with a Lorentz-scale invariant metric ( an invariant proper-time interval in the 
extended target spacetime),  as a result of the embedding, the induced world sheet metric, 
$H_{ab}$ will automatically be scale invariant because under scalings of $resolutions$ the coordinates, $\sigma^a$ are inert. Since  : 
$H_{ab}d\sigma^a d\sigma^b= {\cal G}_{MN}dZ^M dZ^N$,  and the latter interval   is scale-relativistic invariant by construction, it 
follows that $\delta_\beta   (H_{ab}d\sigma^a d\sigma^b)= [\delta_\beta   H_{ab}]d\sigma^a d\sigma^b =0$. Since this is true for all displacements $d\sigma^a$ then $\delta_\beta  H_{ab}=0$. Therefore, the  world sheet area element must be invariant as well because each component of the two-dim metric , $H_{ab}$ is invariant under scale-relativistic transformations. 
Let check that this is true. The metric ${\cal G}_{MN}(Z^M) $ and the $Z^M$ obey  the following :

$$\delta_\beta  H_{ab} =[\delta_\beta {\cal G}_{MN}]\partial_a Z^M \partial _b Z^N
+{\cal G}_{MN}\partial_a [\delta_\beta  Z^M ]\partial _b Z^N+
{\cal G}_{MN}\partial_a Z^M \partial_b [\delta_\beta  Z^N]=0. \eqno (3.20) $$
if : 
$$\delta _\beta {\cal G}_{MN}=(\partial_{Z^M}  {\cal G}_{MN}) \delta_\beta  Z^M = 
 (\partial_{Z^M}  {\cal G}_{MN}) A(\beta,\delta) Z^M =-2A {\cal G}_{MN}=$$

$$M (\beta,\delta) {\cal G}_{MN} \Rightarrow -2A =M . \eqno (3.21)$$
and 

$$  \delta _\beta {\cal G}_{MN}=M (\beta,\delta) {\cal G}_{MN}.~~~\delta_\beta  Z^M= A (\beta,\delta)  Z^M. \eqno (3.22) $$

From eqs-( 2.2, 2.11) we learnt that : 
$$\delta_\beta (\delta) =(\Delta \beta) \beta \gamma^2 \delta \Rightarrow 
\delta_\beta Z^D=(\Delta \beta) \beta \gamma^2 Z^D.$$

$$ \delta_\beta (\varphi/\varphi_o) =-(\Delta \beta) (\gamma^2 \delta)(\varphi/\varphi_o) \Rightarrow 
\delta_\beta Z^m=-(\Delta \beta) (\gamma^2 \delta) Z^m. \eqno (3.23)$$
since :
$$  {\cal G}_{Z^D Z^D}=1   \Rightarrow     \partial_{Z^M}  {\cal G}_{Z^D Z^D}=0 \Rightarrow \delta_\beta   {\cal G}_{Z^D Z^D}=0.~~~ \partial_a Z^D=0.$$

$$ 
\delta_\beta  Z^m=-(\Delta \beta) (\gamma^2 \delta) Z^m.~~~\partial_a (\delta_\beta  Z^m) =
- (\Delta \beta) (\gamma^2 \delta) 
\partial_a Z^m. $$
because the scaling dimension only depends on the resolutions $\Delta A$ 
and  :
$$  {\cal G}_{mn}=-{\delta _{mn} \over \sum (Z^l)^2} \Rightarrow 
\partial_{Z^m}   {\cal G}_{mn}={2Z^m \over [\sum (Z^l)^2]^2}
\Rightarrow \sum Z^m\partial_{Z^m}  {\cal G}_{mn}=-2{\cal G} _{mn}.\eqno (3.23)
$$
with $m=i=0,1,2.....D-1$.
Using the equations above it is straightforward to show that $\delta_\beta  H_{AB}=0$ due to relationship $2A+M=0$ and $\delta_\beta {\cal G}_{mn} =-2A  {\cal G}_{mn}$,  
without introducing any constraints whatsoever
on the 
$Z^M$ variables because the $(\partial_a Z^M) (\partial_a Z^N)$ terms $decouple$ ( can be factored out ). 
This problem was raised earlier in [3]. And conversely, if   $\delta_\beta  H_{AB}=0$ one can show that the 
metric 
${\cal G}_{mn}, {\cal G}_{Z^DZ^D}$ components have the required form as in eq-(3.15) if one does not wish to 
constrain the variables $Z^M$.

Hence one arrives at :

$$Z^D( \Delta A)=T(\Delta A) \leftrightarrow \delta ( \Delta A).~~~
X^\mu \leftrightarrow 
\int  {d\varphi^i (\sigma^a,  \Delta A) \over \sqrt { \sum_i (\varphi^i( \sigma^a, \Delta A ))^2}}. \eqno (3.25)$$
and conclude that Dirac-Nambu-Goto actions are scale-relativistic invariant if, only if, one embeds the string in 
$D+1$ dimensions. The same argument applies to 
all $p$-branes/extendons , gauge invariance of actions under volume-resolutions scaling-relativistic transformations associated with the fuzzy world-volume in  a $D+1$ spacetime with the extra scaling-temporal dimension .

\bigskip

\centerline{\bf IV Concluding Remarks  }

\bigskip

Summarizing, we have shown that  area-resolutions   scale-relativistic invariance is a symmetry of string theory  
 that requires embedding the  $D$ coordinates  $X^\mu$ into $D+1$ dimensions with the extra temporal variable 
being precisely the common scaling dimension of all the string coordinates w.r.t scale-relativistic transformations. In one scoop we have achieved :

1) Why strings cannot probe distances below the Planck scale.

2) Why the string coordinates behave like matrices from the ``fuzzy''  world sheet point 

of view.

3) Why extra temporal dimensions appear in strings.

4) All extended objects admit similar symmetries when areas are replaced by volumes. 

Therefore, string theory,  membranes, etc....are all unified in this manner.

It remains to study the string propagation in curved backgrounds from the $scalings$  point of view and to write the scale 
relativity analog of the the Callan-Symanzik Renormalization Group Equation associated with a 
whole family of actions that respect scale-relativistic invariance. In eqs-(3.20-3.22) we saw that the 
form of the metric ( up to diffeomorphisms)  is tightly 
constrained as a result of scale-relativistic invariance/covariance. 
``Scaled-curved'' metrics must deviate from the scale-flat form in (3.15); i.e the scale-relativity 
version of Einstein's equations obey equations of the  
Callan-Symanzik Renormalization Group type. This was also noticed by Nottale [2]. 

It is unknown why the string quantum effective actions give the classical 
background Einstein,  Yang-Mills, antisymmetric tensor , dilaton ,.....equations of motion . 
Strings can consistently propagate in those ( conformal)  backgrounds if , and only if, the $2D$ 
CFT beta functions 
associated with the string couplings to the background fields vanishes. A sort of quantum/classical duality seems to be operating from the world-sheet/spacetime view. The idea that a quantum/classical duality might exist in Nature has been previously discussed by Nottale pertaining small/large scales : Quantum like structures emerge at the very large and the very small scales. The classical physics regime lives in between. This is another manifestation of the analog  of the 
$T$ duality symmetry in string theory operating in scale-relativity.

The fact that scale-relativity 
invariance, at the classical level , already constrains the form of the background metric is a very promising 
fact that we believe may  answer 
why there is a connection between the quantum string effective action and the classical background field equations. A very important work 
concerning universality  and integrability has been provided  by Fairlie et al [32]. An infinity of Lagrangians furnished the same universal equations of motion. 
Is scale-relativistic invariance tantamount of universality ? in the sense that nonperturbative string  physics exhibits duality symmetries among different Lagrangians that describe the same theory in different corners of the moduli space ? . In the same fashion that ordinary relativistic covariance is tantamount 
of the independence  on the coordinate systems to describe physical phenomena, scale-relativity might signal 
the independence of nonperturbative string physics on the  redundant Lagrangian descriptions .     

Area-momentum uncertainty
relations of a string based on their propagation  in loop spaces have been recently analyzed by [13, 21] where they studied the Hausdorff dimension and fractal like behaviour of the string's world sheet. The area played the role of temporal evolution parameter. The scale relativity modifications of such area-momentum uncertainty are straightforward following our results in [3] based on [2]. Pavsic [20] also  has discussed the propagation of strings and $p$-branes from a loop space point of view and 
wrote wave-functional equations of motion of the 
Wheeler-De Witt type ( Schrodinger like) . 

The fact that points really do not exist as such due to their  smearing and fuzzy-like behaviour into string-bits, 
area-bits, volume-bits,  might bear important connection with the work of [24] and with 
Quantum Groups [25]  and Connes Noncommutative  Geometry. 
The latter made its first appearance in strings with 
Witten open string field theory formulation using BRST 
and path integral techniques which culminated with Zweibach   closed string field theory based on 
Batalin-Vilkovski Antibracket  algebraic (operads)  methods. Is the circle finally closed ? 

The extensions of ordinary $2D$ conformal field theories, $W_\infty $ CFT and $W_\infty$ geometry , 
deserves further study than  the performed so far. What is $W_\infty$ geometry ? , does 
scale-relativity provide clues to find an  answer ? . The connection among $W_\infty$ noncritical strings, affine Toda solitons, integrable models, continuous Toda theories, self-dual membrane, $SU(\infty)$ self dual Yang Mills, Plebanski's heavenly equations, quantum Lie algebras, Moyal deformation
quantization, etc   was provided by the author in [29] based on earlier work by Chapline and Yamagashi,  
Nissimov and Pacheva and many others . We refer to [29] for an extensive  list of references.

Physical applications of Finsler geometries in connection to the $maximal$  proper four-accelerations in string theory ( minimal scale ) have been
discussed by Brandt 
[8]. Conformal Weyl-Finsler structures  has been studied by 
[9]. These Finsler metrics are $no$ mathematical curiosities : these metrics are $imposed$ by Stringy-Physics : 
``maximal'' proper four accelerations. Weyl-Finsler Geometry is thus another natural and plausible 
geometrical setting to start  ( and attempt) to build  the geometrical foundations of string theory. 
Weyl-Finsler geometries allow for the introduction of Torsion as well. Riemannian geometry is recovered in a certain limit. In particular,  Einstein's equations appear in the limit of infinite maximal proper acceleration by taking the  $\Lambda \rightarrow 0$ limit, the analog of the $c\rightarrow \infty$ limit  : the Galilean limit.

A challenging question would be if one can maintain scale-relativity invariance at the $quantum$ level. 
The construction of General Scale-Motion Relativity remains open. We offered some clues at the end of {\bf 3.2}. 
At the quantum level fractals should  have a pivotal role. To go beyond the 
Quantum String Geometry [33] may require generalizations of Riemannian geometry 
that include curvilinear fractal coordinate systems [2]. A notion of fractal derivative, 
fractal integration, fractal measure, .....appearing in fractal geometries  has already been built. What remains is to 
construct the version  of metrics, connections, curvatures....that would enable us to define 
inertial and accelerated systems of reference in fractal spacetimes.  
We are unaware if such mathematical tools are avalilable to-day and for this reason we have  
followed  the simplest route that has been paved over the years and that  originated with  Weyl : 
gauge invariance. Hints that p-adics numbers and non Archimedean geometries  might very relevant to tackle this very difficult challenge at the Planck scale have been given among many others in [28].

\bigskip
\centerline {\bf Acknowledegements}

We are greatfuly indebted to Laurent Nottale for discussions and to the Center for Theoretical Studies of Physical Systems for support.  

\bigskip

\centerline {\bf REFERENCES}

1-. A. Tsetlyn : Int. Jour. Mod. Phys. {\bf A 4} (1989) 1257

2-. L. Nottale : `` Fractal Space Time and Microphysics : Towards the Theory

of Scale Relativity''. World Scientific. 1992

L. Nottale :Int. Jour, Mod. Phys. {\bf A 4} (1989) 5047.  Int. Jour. Mod. Phys. {\bf A 7} (1992) 4899.

L. Nottale : `` Scale Relativity'' in  `` Quantum Mechanics, Diffusion and Chaotic 

Fractals, vol II. Pergamon Press , 1996. 

L. Nottale : `` Scale Relativity and Quantization of the Universe : 

Theoretical Framework `` to appear in the J. of Astronomy and Astrophysics, (1997).

G. Ord : Journ. Physics A. Math General {\bf 16} (1983) 1869.

3-. C. Castro : `` String Theory, Scale Relativity and the Generalized 

Uncertainty Principle `` to appear in Foundations of Physics Letters {\bf 10} (3) (1977). 

4-. D. Amati, M. Ciafaloni, G. Veneziano : Phys. Lett {\bf B 197} (1987) 81.

D. Gross, P. Mende : Phys. Letters. {\bf B 197} (1987) 129.

5-. L. Garay : Int. Journ. Mod. Phys {\bf A 10} (1995) 145.

6-. A. Ahtekar, C. Rovelli, L. Smolin : Phys. Rev. Letters {\bf 69} (1992) 237.

7-. Y. Ne'eman, D. Sijacki : Mod. Phys. Lett {\bf A 11} (1996) 217.

8-. H. Brandt : `` Finslerian Space Time `` Contemporary Mathematics Series 

of The American Mathematical Society. {\bf 196} (1996) 273.

9-. T. Aikou, Y. Ichijyo : Rep. Fac. Science, Kagoshima Univ. {\bf 23} 

(1990) 101.

10-. J. Ellis, N. Mavromatos, D. Nanopoulos : Phys. Lett.  {\bf B 293 } (1992)

37. Mod. Phys. Lett {\bf A 10} (1995) 425. 

J. Ellis, N. Mavromatos, D. Nanopoulos : `` $D$ Branes from Liouville Strings `` 

hep-th/9605046

F. Lizzi, N. Mavromatos : `` Quantum Phase Space From String Solitons ``

hep-th/9611040

11-. E.Bergshoeff, E.Sezgin, Y.Tanni, P.Townsend : Annals of Phys. {\bf 199}

(1990) 340.

E. Guendelman, E. Nissimov, S. Pacheva : `` Volume-Preserving Diffeomorphisms

versus Local Gauge Symmetry''. hep-th/9505128.  

A. Aurilia, A. Smailagic and E. Spallucci : Phys. Rev {\bf D 47} (1993) 2536.

C. Castro : `` p-Branes as Composite Antisymmetric Tensor Field Theories'' 

to appear in the Int. Journal of Mod. Physics A.

12-M. Duff : Class. Quant. Gravity {\bf 6} (1989) 1577.

13- S. Ansoldi, A. Aurilia,  and E. Spallucci : `` Hausdorff dimension of the Quantum 

String `` hep-th/9705010.

14-A. Kempf, G. Mangano : `` Minimal Length Uncertainty and Ultraviolet 

Regularisation ``  hep-th/9612084. 

G. Mangano : `` Path Integrals in Noncommutative Spaces `` gr-qc/9705040.

15- T. Thiemann : `` QSD V : Quantum Gravity as a Natural Regulator of Matter 

QFT ``  gr-qc/9705019.

16- L. Smolin : `` The Future of Spin Networks ``  gr-qc/9702030.

17-T. Nowotny, M. Requard : `` Dimension Theory of Graphs and Networks `` 

hep-th/9707082.

18-D. Datta : `` Duality and Scalings in Quantum Mechanics ``    hep-th/9707055

19-J. Madore : `` Classical Gravity in Fuzzy Spacetime `` gr-qc/9611026.

20- M. Pavsic : `` The Dirac-Nambu-Goto p-branes as Particular Solutions 

to a generalized Unconstrained Theory ``IJS-TP-96-10 preprint.

21- S. Ansoldi, A. Aurilia,  and E. Spallucci : `` String Propagator, A Loop Space 

Representation `` hep-th/9510133.

22- M. Cederwall, G. Ferreti, B. Nilsson, A. Westerby :  

Nucl. Phys. {\bf B 424} (1994) 97.

23- T. Yoneya : `` D-particles, D- Instantons, and a Space Time Uncertainty Principle 

in String Theory ``  hep-th/9707002.

24- H. Noyes : `` An Introduction to Bit String Physics ``  hep-th/9707020.

C. Thorn : `` Supersymmetric Quantum Mechanics for String Bits `` hep-th/9707048.

25- G. Amelino-Camelia, J. Lukiersky, A. Nowicki : `` $\kappa$ deformed covariant phase 

space and quantum gravity uncertainty relations `` hep-th/9706031.

26- E. Witten : Nucl. Phys. {\bf B 460} (1996) 333.

J. Polchinski : `` Tasi Lectures on $D$-branes'' hep-th/9611050. 

T. Banks, W. Fischler, S. Shenker and L. Susskind : `` M theory as a Matrix model,

a conjecture ``   hep-th/9610043.

J. Hoppe : ``Quantum Theory of a Relativistic Surface `` MIT, Ph.D thesis. 1980. 

R. Flume : Ann. Phys. {\bf 164} (1985) 189.

M. Baake, P. Reinicke, V. Rittenberg : J. Math. Phys. {\bf 265} (1985) 1070. 

27- A. Connes : `` Non Commutative Geometry `` Academic Press. New York 1994.

28- V. Vladimorov, I. Volovich, E. Zelenov : `` p-adic analysis and Mathematical 

physics `` World Scientific, Singapore, 1992.

29- C. Castro : `` A Moyal Quantization of the Continuous Toda Field `` to appear in 

Physics Letters B.

G. Chapline, K. Yamagishi : Class. Quant. Grav. {\bf 8} (1991) 427.

E. Nissimov, S. Pacheva : Theor. Math. Phys. {\bf 93} (1992). 

30- I. Bars : `` A case for 14 Dimensions '' hep-th/9704054 

31- C. Vafa : `` Evidence for F Theory `` hep-th/9602022

32- D. Fairlie, J. Goaverts, A. Morozov : Nucl. Phys {\bf B 373} (1992) 214.

33- B. Greene : `` String Theory on Calabi Yau Manifolds `` hep-th/9702155.

P. Aspinwall : `` K3 Surfaces and String Duality `` hep-th/9611137.

\bye